\documentclass{iopart}
\usepackage{iopams}
\usepackage{epsfig}
\usepackage{epstopdf}

\begin{document}


\title{Entanglement and transport anomalies in nanowires}
\author{ J.H. Jefferson$^a$, A. Ram\v sak$^b$, and T. Rejec$^b$}
\address{$^a$QinetiQ, Sensors and Electronics Division, St. Andrews Road,
Great Malvern, England\\
$^b$Faculty of Mathematics and Physics, University of Ljubljana, and
Jo\v zef Stefan Institute, Ljubljana, Slovenia
}

\begin{abstract}
A shallow potential well in a near-perfect quantum wire will bind
a single-electron and behave like a quantum dot, giving rise to
spin-dependent resonances of propagating electrons due to Coulomb
repulsion and Pauli blocking. It is shown how this may be used to
generate full entanglement between static and flying spin-qubits
near resonance in a two-electron system via singlet or triplet
spin-filtering. In a quantum wire with many electrons, the same
pairwise scattering may be used to explain  conductance,
thermopower and shot-noise anomalies, provided the
temperature/energy scale is sufficiently high for Kondo-like
many-body effects to be negligible.
\end{abstract}

\noindent

\section{Introduction}

The scattering of propagating electrons from a single bound electron
in a hydrogen atom was first solved over 75 years ago by
Oppenheimer \cite{oppenheimer} and Mott \cite{mott}
who showed that, due to the indistinguishability of
electrons and the mutual Coulomb repulsion between the propagating
electron and the bound electron, the scattering was strongly dependent
on total spin, with resonances at different energies for singlet and
triplet configurations. With the remarkable progress in semiconductor
device fabrication, quantum wire and quantum dot structures may now be
fabricated which enable the scattering of propagating electrons from a
single bound electron to be studied. We can regard such a system as a
one-dimensional analogue of the H-electron scattering system but at a
much lower energy scale.  In this paper we shall consider such
mesoscopic systems and their potential for studying spin-dependent
scattering effects including how they differ from the
three-dimensional case, how spin-dependent scattering might be
utilised, and review the experimental evidence for the predicted
effects.

We shall consider device structures which are near perfect in the
sense that transport between source and drain contacts is coherent and
the effects of electron-electron scattering is not masked by strong
disorder scattering or incoherent scattering due to phonons. This may
be achieved in practise with high-quality gated two-dimensional
electron sheets (2DES) at low bias and sufficiently low temperatures
in which only the lowest subband, or channel, is occupied. The
technology of such devices based on GaAs has reached a high level of
maturity following the pioneering work on conductance quantisation in
point contacts \cite{wees88, wharam88} through to gated
quantum-dot structures in which single-electrons may be captured and
moved from one location to another utilising time-dependent gate
potentials \cite{koppens, petta} or the propagating confinement potential provided by surface
acoustic waves \cite{kataoka}. Although we shall consider explicit
examples based on such 'soft-confined' structures the effects that we
describe are generic and may in principle also be observed in
hard-confined structures such as etched or directly grown quantum
wires \cite{yacoby96, kristensen98,franceschi,jespersen} or molecular structures,
semiconducting carbon nanotubes for example, in which electrons may be
either confined by gates \cite{jorgensen} or by fullerenes inside the nanotube, the
so-called peapod structures \cite{benjamin}. The potential
advantage of such hard-confined devices is that the increased
confinement, with sub-nanometer length scales in two or even all three
dimensions, give corresponding enhanced energy scales with the
possibility of observing at least some of the effects at elevated
temperatures, up to room temperature. However, with the present state
of the technology, we are still some way from controlled fabrication of
clean devices with good ohmic contacts.

The paper is organised as follows. In the next section we shall
consider the problem of the scattering of a propagating electron
from a bound electron in a quantum wire and show how this may be
utilised to generate entanglement. In the next section, the
results are generalised to the case of many electrons in the wire,
scattering from a single bound electron. We then show how this
gives rise to conductance anomalies, the so-called 0.7 and 0.3
anomalies, as a consequence of spin-dependent scattering. The
basic solutions of the scattering problem are then used to explain
anomalies in thermopower, thermal conductance and shot-noise. We
conclude with a discussion of many-body effects expected at
low-temperatures, particularly the Kondo effect, and how they may
be incorporated into the basic framework based on spin-dependent
two-electron scattering.

\section{Two-electron problem and entanglement}

\begin{figure}[hbt]
\center{\includegraphics[width=50mm]{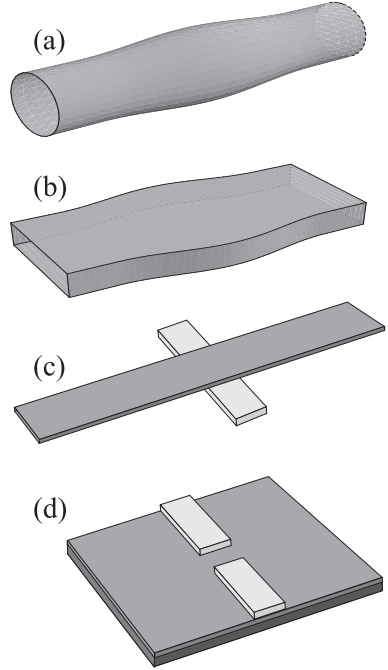}}
\caption{\label{Fig1} Schematic of various realisations of a near-perfect
quantum wire. (a ) A cylindrical and (b) a rectangular wire with a
bulge. (c) A gated straight wire. (d) Quantum point contact.}
\end{figure}

In this section we establish a generic model for a one-dimensional
quantum wire containing a small region where a single electron may be
bound and consider the problem of the scattering of a second electron
propagating in the wire. At one extreme, this may be a near perfect
quantum wire with a very shallow confining potential which, for
example, might occur within a point contact constriction in a 2DES in
a semiconductor. As we have suggested before
\cite{rrj00,jrr,rejecand}, the origin of the confining potential may
be some fluctuation due to remote defects or gates or, in very clean
systems, may be due to the single electron itself, a possibility that
has received some recent support both theoretically
\cite{rejecmeir,sushkov} and experimentally \cite{yoon}. The potential
well may also be created (or enhanced) with a narrow strip-gate
perpendicular to a quantum wire such as in a carbon nanotube
\cite{gunlycke}. Examples are shown schematically in
Fig.~\ref{Fig1}. We may refer to this confined region as an open
quantum dot since it is not defined by explicit barriers, the
restriction to one bound electron being enforced purely by the Coulomb
interaction.  We do not need to know the explicit cause or the details
of this weak confining potential, the only requirement is that it is
sufficiently weak to allow only one bound electron.  In the opposite
extreme we may have a well-defined quantum dot with explicit barriers
as, for example, produced in 2DES devices \cite{koppens, petta} and
carbon nanotubes \cite{jorgensen}. A further possibility is that the
quantum dot is actually a physically separate entity from the quantum
wire, such as would occur in a carbon-nanotube peapod in which one or
more fullerenes inside the nanotube contain a single electron in their
highest occupied molecular orbitals. It is well known that the
current-voltage characteristics of devices fabricated from such
systems can be quite different. Near perfect quantum wires show
conductance steps at multiples of $2e^2/h$, whereas quantum-dot based
devices give rise to Coulomb blockade and the characteristic 'diamond'
plots of conductance as gate and source-drain voltages are varied.  In
this paper we argue that in the narrow regions of parameter space
where these systems begin to conduct they have similar behaviour which
is a consequence of spin-dependent scattering enforced by the mutual
Coulomb repulsion between two electrons in the quantum-dot region,
coupled with Pauli exclusion when the spins are parallel.

A general  Hamiltonian for the two-electron problem has the form
\begin{equation}
H=-\frac{\hbar^{2}}{2m}\left[\frac{\partial^{2}}{\partial
x_{1}^{2}}+\frac{\partial^{2}}{\partial x_{2}^{2}}\right]+
v(x_{1})+v(x_{2})+V(x_{1},x_{2}).
\label{hamilt2}
\end{equation}
In this equation we assume sufficiently high confinement that only
the lowest non-degenerate transverse mode (channel) is occupied
and the two electrons move in the effective one-electron
potential, $v(x)$, with effective-mass $m$. This is not a serious
restriction and corrections due to occupation of higher transverse
mode, either due to thermal excitation or inter-channel coupling,
may readily be accounted for perturbatively. The analysis and
results may be extended to cases where there is degeneracy but we
will not consider them here since our main aim is to emphasise the
generic effects in the simplest and most direct way.  The
one-electron potential accounts for potential fluctuations due to
disorder or smooth potential changes due to external gates, for
example in a quantum-dot structure where they would model
barriers. We shall be particularly interested in near-perfect
wires for which $v(x)$ represents a shallow potential well (open
quantum dot) that will bind a single electron, but not two
electrons. The Coulomb interaction between the two electrons,
$V(x_1,x_2)$, may be derived explicitly by starting with the usual
3D form and integrating over the lowest occupied transverse mode.
The resulting expression, to a very good approximation, takes the
simplified form $V(x_1,x_2)=e^2/\epsilon r$, where
$r=\sqrt{(x_1-x_2)^2+\lambda^2}$ is the effective separation and
$\lambda$ is a parameter that characterised the transverse
confinement, e.g., $\lambda \sim d$ for a circular wire of
diameter $d$ \cite{rrj00}.

We now consider the two-electron scattering problem in which a
single electron is first introduced into the quantum wire and
becomes bound in the quantum-dot region. The confining potential
for this electron is determined completely by the one-electron
potential $v(x)$.  To be specific, we shall mainly consider a
quantum wire with material parameters appropriate to GaAs, i.e.
$m=0.067m_{0}$ and $\epsilon=12.5$, and a potential profile with
$v(x)=-v_0\cos ^{2}\pi x/L$ for $-L/2<x<L/2$ and zero otherwise.
One motivation for studying this problem is the rapidly emerging
field of quantum information processing in which a major goal is
the generation of entanglement the controlled exchange of quantum
information between propagating and static qubits. Purely electron
systems have potential as entanglers due to strong Coulomb
interactions and although charge-qubit systems suffer from short
coherence times, spins in semiconductor quantum wires and dots are
sufficiently long-lived for spin-qubits to be promising candidate
for realising quantum gates involving both static and propagating
spins~\cite{Loss98,Elzerman04,Hanson03,Hanson05,Petta04}. Electron
entanglers have been proposed using a double-dot, exploiting the
singlet ground state~\cite{hu04}; the exchange interaction between
conduction electrons in a single dot~\cite{costa01,oliver02}; and
the exchange interaction between electron spins in parallel
surface acoustic wave channels~\cite{Barnes00}.  Measurement of
entanglement between propagating electron pairs has also been
proposed using an electron beamsplitter~\cite{burkard04}.  A
further motivation for solving the two-electron scattering problem
is that the solutions may be used directly in expressions for
conductance and other transport quantities, which we consider in
the next section.

Before presenting detailed numerical results, consider the expected
general features when a spin-up propagating electron interacts with a
bound electron in some arbitrary state on the Bloch sphere,
i.e. $\cos(\vartheta/2)\left|
\downarrow\right>+e^{i\phi}\sin(\vartheta/2)\left|\uparrow\right>$. In
practise this situation may be realised, at least in principle, by a
system in which single spin-polarised electrons are introduced into
the wire through a turnstile injector. This first electron is then
allowed to relax into the ground-state of the quantum dot region and
its spin is subsequently manipulated on the Bloch sphere using
microwave pulses. The second spin-up electron then scatters from this
bound electron. 
It is important to realise that this two electron state before scattering is 
unentangled and is not an eigenstate of the Hamiltoninan Eq.~(\ref{hamilt2}).
It may be written,
\begin{equation}
\Psi_\mathrm{in}=\cos(\theta/2)\Psi_{\uparrow\downarrow}+
e^{i \phi}\sin(\theta/2)\Psi_{\uparrow\uparrow},\label{psiin}
\end{equation}
where
\begin{equation}
\Psi_{\sigma \sigma'}=e^{i k x_1} \psi_0(x_2)\chi_\sigma(\sigma_1)
\chi_{\sigma'}(\sigma_2)-
e^{i k x_2} \psi_0(x_1)\chi_\sigma(\sigma_2)
\chi_{\sigma'}(\sigma_1). \label{psissp}
\end{equation}
Here $\psi_0$ is the bound-state wavefunction for the quantum-dot region
and $\chi_\sigma(\sigma_{1,2})$ is a spin-1/2 
spinor for an electron with spin $\sigma_{1,2}$.

During the scattering process the two electrons interact resulting in
reflection or transmission.  The scattered component resulting from
$\Psi_{\uparrow\downarrow}$ in Eq.~(\ref{psiin}) can be either spin-flip or
non-spin-flip scattered with corresponding amplitudes $r_\mathrm{sf}$,
$r_\mathrm{nsf}$, $t_\mathrm{sf}$ and $t_\mathrm{nsf}$ for reflection
and transmission.  Spin-flip scattering is due to an exchange
interaction between bound and propagating electrons.  This is not a
direct interaction between spins of the electrons since the
Hamiltonian Eq.~(\ref{hamilt2}) does not contain any explicit spin-spin
terms.  It arises purely because of the antisymmetry of the
two-electron wavefunction and their mutual Coulomb repulsion. The
transmitted component is
\begin{equation}
\Psi_\mathrm{trans}=\cos(\theta/2)[t_\mathrm{nsf}\Psi_{\uparrow\downarrow}+
t_\mathrm{sf}\Psi_{\downarrow\uparrow}]+
e^{i \phi}\sin(\theta/2)\Psi_{\uparrow\uparrow}
\label{trans}
\end{equation}
with a similar expression for reflection. 

After scattering, the propagating electron will be reflected or
transmitted and, asymptotically, will have the same magnitude of
momentum, $k$, leaving the bound electron again in its ground state,
provided the kinetic energy of the incoming electron is chosen to be
less than the binding energy of the bound electron. It is convenient
and instructive to express these amplitudes in terms of eigenstates of
the Hamiltonian, which have definite total spin since
$[H,S^2]=[H,S^z]=0$.  For 2 electrons the eigenstates are either
singlets or triplets with $\Psi_{00}=(\Psi_{\uparrow\downarrow} -
\Psi_{\downarrow\uparrow})/\sqrt{2}$,
$\Psi_{10}=(\Psi_{\uparrow\downarrow} +
\Psi_{\downarrow\uparrow})/\sqrt{2}$,
$\Psi_{11}=\Psi_{\uparrow\uparrow}$, and
$\Psi_{1,-1}=\Psi_{\downarrow\downarrow}$.  Substituting these
expressions into Eqs.~(\ref{psiin}), (\ref{psissp}) and (\ref{trans})
we see directly that,
\begin{equation}
t_\mathrm{nsf}=\frac{t_{1}+t_{0}}{2}\:\textrm{and}\:
t_\mathrm{sf}=\frac{t_{1}-t_{0}}{2},\label{eq:tnsf}
\end{equation}
where $t_0$ and $t_1$  are the singlet and triplet 
transmission amplitudes, respectively.
Both the reflected and transmitted waves show spin entanglement after
scattering provided $\cos(\vartheta/2)$ and the amplitudes for
spin-flip and non-spin-flip scattering, $r_\mathrm{sf},\,
r_\mathrm{nsf},\, t_\mathrm{sf}$ and $t_\mathrm{nsf}$ are
non-zero. Furthermore, fully entangled states occur when $\vartheta=0$
and $|r_\mathrm{sf}|=|r_\mathrm{nsf}|$ or
$|t_\mathrm{sf}|=|t_\mathrm{nsf}|$. It follows directly from 
this that one way of achieving maximum entanglement 
would be via either a singlet or triplet resonance.  More precisely, if 
either $|t_0|=1$, $|t_1|=0$, or $|t_1|=1$, $|t_0| =0$ then the transmitted state 
will correspond to full entanglement between the transmitted electron and 
the electron remaining in the quantum dot.  This has a simple physical 
interpretation. The initial (unentangled) state is the sum of a singlet 
state and an $S^z=0$ triplet state and only one of these components is 
transmitted or 'filtered'.  Since either a singlet or a triplet state is 
fully entangled, such filtering via resonance produces a fully entangled 
static-flying spin-qubit pair with probability 1/2. Furthermore, the 
reflected state would also be fully entangled but with the complementary 
component.  In this way the initial unentangled state is divided into its 
fully entangled singlet and triplet components, separated by transmission 
and reflection.

In practise it is not immediately obvious that such resonances will occur in 
quantum wires and in any case, the transmission of the complementary 
component is not expected to be precisely zero so the question arises as to 
whether perfect entanglement can still be achieved.   In fact, the 
two-electron system in which there is a shallow quantum well will always 
have at least a singlet resonance.

\begin{figure}[hbt]
\center{\includegraphics[width=60mm]{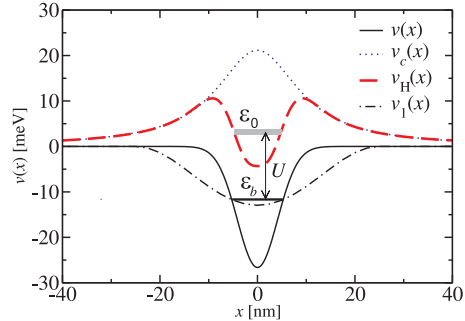}}
\caption{\label{Fig2}A shallow potential well $v(x)$ (full line) and
the corresponding Hartree potential seen by a second electron of opposite
spin (dashed line). The single resonant bound state, $\epsilon_{b}$,
and quasi-bound singlet state, $\epsilon_0$, within the double
barrier structure are also indicated. The dotted line is the Coulomb
repulsion energy due to bound electron and the dashed-dotted line
represents shallow potential well $v_{1}(x)$ corresponding for the
results from Fig.~\ref{Fig3}(b).}
\end{figure}

This two-electron system has at least a singlet resonance
($|t_{0}|=1)$ at some energy for which the triplet state is off
resonance $(|t_{1}|\ll1)$, as may be seen from explicit solutions of
the scattering problem.
We may choose the shallow effective potential, $v(x)$, such that
there is only a single one-electron bound state. For example, with a well
depth of $26$~meV and a width of
$\sim20$~nm (Fig.~\ref{Fig2}) there is a single bound state at energy
$\epsilon_{b}=-12$~meV. The bound electron has a long-range
Coulomb interaction with the propagating electron and, when combined
with the well potential, gives rise to a double barrier structure
which has a singlet resonance energy at approximately
$\epsilon_0\sim\epsilon_{b}+U$, where $\epsilon_{b}$ is the energy of
the lowest bound state and $U=15$~meV is the Coulomb matrix element
for two electrons of opposite spin occupying this state. This is also
shown in Fig.~\ref{Fig2} where we have plotted the Hartree potential
due to the bound electron, $v_\mathrm{H}(x)$, i.e. the self-consistent
potential seen by the propagating electron in the 'frozen' potential
due to the bound electron of opposite spin.
For wider wells a second
bound state is allowed and this will give rise to a triplet resonance
(and a further singlet resonance) at higher energy. The
singlet-triplet separation may be controlled by changing the width and
depth of the well whilst maintaining the condition that only one
electron be bound. With increasing well-width (and decreasing
well-depth) the singlet-triplet separation reduces and eventually the
resonances overlap.

At the lowest (singlet) resonance,
$|t_\mathrm{sf}|\approx|t_\mathrm{nsf}|\approx\frac{1}{2}$ and the
state is close to being fully entangled when spins are initially
antiparallel ($\cos(\vartheta/2)=1)$.  Similarly, in reflection,
$|r_{0}|\approx0$, $|r_{1}|\approx1$ and
$|r_\mathrm{sf}|\approx|r_\mathrm{nsf}|\approx\frac{1}{2}$, which
is also close to being fully entangled. The precise condition for full
entanglement in transmission, when the triplet component is not
precisely zero is $\vartheta=0,$ $|t_{1}+t_{0}|=|t_{1}-t_{0}|$.
This may be seen directly from the form of asymptotic outgoing
states. When the scattered components are not fully entangled, we need a 
quantitative measure of entanglement which reduces to unity for fully 
entangled states and zero for unentangled states.  Such a measure is the 
concurrence which, for the pure states considered here, takes the form 
Ref.~\cite{wootters98}
\begin{equation}
C=2|\langle \Psi | \Psi_{\downarrow\downarrow}\rangle
\langle \Psi | \Psi_{\uparrow\uparrow}\rangle-
\langle \Psi | \Psi_{\uparrow\downarrow}\rangle
\langle \Psi | \Psi_{\downarrow\uparrow}\rangle|,
\label{c2}
\end{equation}
where $\Psi$ is the scattered component in either transmission or reflection. 
We can see immediately from this expression that $C=1$ for fully entangled 
states, such as a singlet or $S^z=0$ triplet state, and zero for an unentangled 
state, such as $\Psi=\Psi_{\uparrow\downarrow}$.  Substituting 
Eq.~(\ref{trans}) into Eq.~(\ref{c2}), the concurrence becomes
\begin{equation}
C=\frac{2\cos^2(\vartheta/2)|t_\mathrm{sf}\, t_\mathrm{nsf}|}
{\cos^2(\vartheta/2)\left[|t_\mathrm{sf}|^2+|t_\mathrm{nsf}|^2\right]+
\sin^2(\vartheta/2) |t_\mathrm{sf}+t_\mathrm{nsf}|^2}.\label{eq:ct}
\end{equation}

\begin{figure}
\center{\includegraphics[width=130mm]{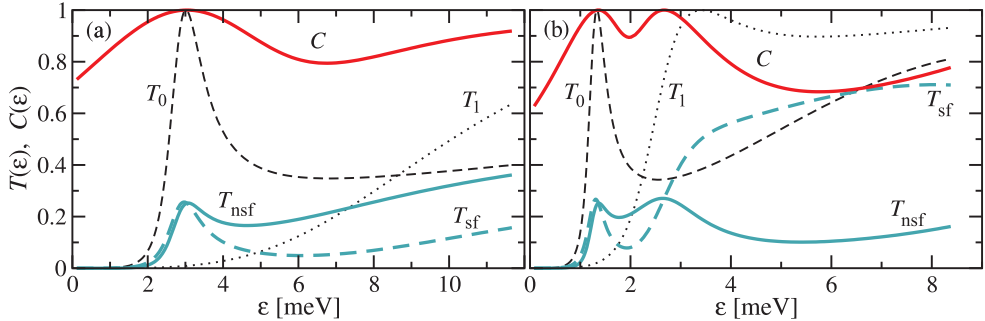}}
\caption{\label{Fig3}(a) Singlet and triplet transmission
probabilities ($T_S=|t_S|^2$), spin-flip and non-spin-flip
transmission probabilities and corresponding concurrence $C$ for
confining potential $v(x)$ from Fig.~\ref{Fig2}. (b) Results for a shallower
and longer potential, corresponding to $v_1(x)$ from Fig.~\ref{Fig2}.}
\end{figure}

Further illustration of this behaviour is seen by solving the
scattering problem explicitly for specific cases.  Numerical solutions
for symmetrised (singlet) or antisymmetrised (triplet) orbital states
yield directly the complex amplitudes $t_{0},\, r_{0},\, t_{1},\,
r_{1}$ from which the amplitudes for spin-flip and non-spin-flip
scattering may be calculated using Eq.~(\ref{eq:tnsf}).

In Fig.~\ref{Fig3}(a) we plot the singlet and triplet transmission and
reflection probabilities, showing a single maximum of unity for the
transmission. We also plot concurrence, which reaches a maximum value
very close to the corresponding singlet resonance, occurring when the
spin-flip probabilities are equal and approximately $\frac{1}{4}$ in
both transmission and reflection.
In Fig.~\ref{Fig3}(b) we show results for a shallow potential well of
depth $12$~meV and width $\sim40$~nm. With these parameters there are
two single-electron bound states at energies $-8$~meV and
$-10$~meV. This gives both a singlet and a triplet resonance, the
latter corresponding to one electron in the lowest bound state and the
other in the higher bound state which becomes a resonance obeying
Hund's rule under Coulomb repulsion, with a further singlet resonance
outside the energy window for elastic scattering. We see that there
are two unitary peaks of concurrence in transmission with the second
close to, but clearly discernible from, the peak of the rather broad
triplet resonance.

Thus we have shown that spin-entanglement and exchange of quantum
information occurs via the Coulomb interaction when a propagating
electron interacts Coulombically with a single bound electron in a
shallow potential well in a one-dimensional semiconducting quantum
wire. The degree of entanglement may be controlled by kinetic energy
of the incoming electron and the shape of the effective potential well
and unitary concurrence occurs near a singlet or triplet resonance.
Potential realisations of such a system are semiconductor quantum
wires and carbon nanotubes.

We conclude this section by describing the differences between the
case of a near perfect quantum wire, or open quantum dot, described
above and a 'real' quantum dot defined by barriers. The presence of
barriers, which in practise may be produced by further gates in 2DES
systems, give the well-known single-electron tunneling resonances
when the kinetic energy of the incoming electron is resonant with a
quasi-bound state in the quantum dot. The first resonance occurs when
there is no energy cost for a single-electron for hop onto or off the
dot and the charge on the dot fluctuates between 0 and 1, with no real
bound state. There is no such single-electron regime in the quantum
wire case for which there is a real bound state that will bind one
electron. In the case of the quantum dot, a second single-electron
tunnel-ling regime occurs when the depth of the potential well is
increased so that one electron is now in a real bound state and a
second electron of opposite spin may resonantly tunnel on and off the
dot at energy $U$ higher than the bound state, where $U$ is the
intra-dot Coulomb matrix element. This also occurs in the quantum wire
case but here the barriers arise solely from the Coulomb repulsion
between the two electrons. One consequence of this is that at higher
energies than the resonance, after an initial dip the transmission
tends smoothly to unity, as shown in Fig.~\ref{Fig3}, whereas in the
dot case the conductance remains low away from resonance. However, in
other respects the two systems behave in a similar way, giving full
entanglement near a two-electron resonance.

\section{Many-electron problem and transport anomalies}

Close to the conduction edge in a near-perfect quantum wire we again
conjecture that a single electron becomes bound in a shallow potential
well, or open quantum dot, along the wire. This then gives rise to
spin-dependent scattering of conduction electrons and since we are
close to the threshold for conduction it is reasonable to assume that
the system will be dominated by sequential pairwise scattering. At
each scattering event, total $S,S_z$ are conserved since, by
assumption, there are no explicit spin-terms in the
Hamiltonian. Hence, when the spins are parallel their direction cannot
change in the scattering process. On the other hand, there will be
both spin-flip or non-spin-flip scattering when the two electrons
initially have opposite spin. This is a consequence of the effective
exchange interaction between the spins, resulting from the Coulomb
interaction and Pauli exclusion. As shown in the previous section,
this is particularly pronounced when we are close to either a singlet
or a triplet resonance for which a transmitted electron (with
transmission probability ~1/2) is fully entangled with the bound
electron left behind, i.e. the scattering process acts as a spin
filter of the initial state with equal probabilities of spin-flip and
non-spin-flip transmission. More generally, if we sum over all
incident electrons near the Fermi energy then, using the pairwise
scattering solutions derived in the previous section, we may derive
expression for transport quantities. We expect anomalies in these
quantities when the Fermi energy is close to either a singlet or a
triplet resonance. For the former, approximately $1/4$ of electrons
are transmitted since half of them have spin parallel to the bound
electron and of the remaining half, with spin antiparallel,
approximately half are transmitted (singlet spin filter). This gives a
total transmission of order $1/4$ when Fermi energy is close to the
singlet resonance energy. On the other hand, close to a triplet
resonance (at higher energy than the singlet resonance), of order
$3/4$ of the electrons are transmitted. In what follows, we argue
that these spin-dependent resonances are the main source of the
transport anomalies observed in near perfect quantum wires.

\subsection{Conductance anomalies}

In the spirit of the Landauer-B{\" u}ttiker approach~\cite{landauer},
we may determine the current in a quantum wire or point contact from
transmission probabilities.  From equations (\ref{trans}) and
(\ref{eq:tnsf}), the total transmission probability for an incident
spin-up electron interacting with a bound spin-down electron is
\begin{equation}
T_{\uparrow\downarrow}=|t_\mathrm{sf}|^2+|t_\mathrm{nsf}|^2=
{1 \over 2} (|t_0|^2+|t_1|^2).
\label{t2}
\end{equation}
This is the same as the transmission probability for a propagating spin-down 
electron interacting with a bound spin-up electron, $T_{\downarrow\uparrow}$. 
When both 
propagating electrons have the same spin the transmission probability is 
that of a triplet component, i.e., $T_{\uparrow\uparrow} = 
T_{\downarrow\downarrow}= |t_0|^2$.  Summing 
over all electrons in source and drain leads, with equal probabilities that 
the bound electron is initially spin-up or spin-down, gives directly the 
genealised Landauer-B{\" u}ttiker formula \cite{rrj00},
\begin{equation}
I=\frac{2e}{h}\int(\frac{1}{4}T_{0}+\frac{3}{4}T_{1})(f_{L}-f_{R})
d\epsilon,\label{eq:lbg}
\end{equation}
where $T_{S}=T_{S}(\epsilon)$ are the energy dependent singlet or triplet
transmission probabilities for $S=0$ and $S=1$, respectively.
$f_{L,R}=\{1+\exp[(\epsilon-\mu_{L,R})/k_{B}T]\}^{-1}$ is the usual
Fermi distribution function corresponding to left and right lead,
respectively, with temperature $T$ and the Boltzmann constant $k_{B}$.

\begin{figure}[htb]
\center{\epsfig{file=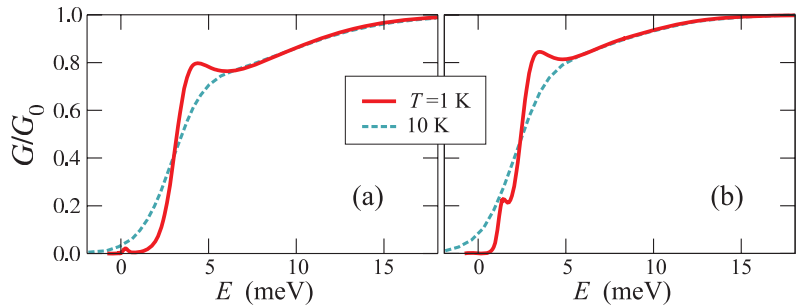,height=40mm,angle=-0}}
\caption{The conductance as calculated from Eq.~(\ref{eq:lbg}) at
  $T=1$~K and $T=10$~K for two different wires with thickness $a=10$~nm: (a)
  $L=40$~nm and $v_0=19$~meV; (b) $L=50$~nm and
  $v_0=13$~meV.
\label{Fig4}
}
\end{figure}

When the wire is connected to metallic source-drain contacts,
electrons will flow into the wire region provided that the Fermi
energy in the contacts is higher than the lowest allowed
eigen-energy in the wire. As the Fermi energy is raised from below
the conduction band edge in the wire (via a gate not considered
explicitly), at least one electron will become bound in the
shallow-well region of the wire. The number of bound electrons
depends on both the Fermi energy and the relative depth and width
of the well. As we have already stated, the parameters are chosen
such that only one electron is bound.  In Fig.~\ref{Fig4} we show
plots of the conductance $G=\mathrm{d}I/\mathrm{dV}$ where
$V=-(\mu_L-\mu_R)/e$ is the voltage difference between the leads,
for two typical wires. Similar results are obtained for wires with
thickness, $a$, up to $a\thicksim 50$~nm, beyond which the
single-channel approximation becomes less reliable as electron
correlations become increasingly important \cite{rrj03}.
The main feature of these results is that there are resonances in both
singlet and triplet channels and these give rise to structures in the
rising edge to the first conductance plateau for $G\sim
\frac{1}{4}G_{0}$ (singlet) and $G\sim \frac{3}{4}G_{0}$ (triplet),
with $G_0=2e^2/h$. The system is therefore in the regime where the
incident electron sees a double barrier which will have some resonant
energy for which there is perfect transmission. A more detailed
analysis has to account for spin and this may be understood by
gradually switching on the Coulomb interaction. For the present choice
of parameters, and also a range of parameter sets which correspond to
a very shallow well, there are two bound states for one electron. With
no interaction both electrons may thus occupy one of 4 states (3
singlets and a triplet).  If we now switch on a small Coulomb
interaction then the lowest two-electron state will be a singlet,
derived from both electrons in the lowest one-electron state. We may
regard one electron as occupying the lowest bound-state level and the
other electron of opposite spin also in this same orbital state but
energy $U$ higher, where $U$ is the intra-`atomic' Coulomb matrix
element, as in the Anderson impurity model. As the Coulomb interaction
is increased, $U$ eventually exceeds the binding energy and this
higher level becomes a virtual bound state giving rise to a resonance
in transmission. An estimate of the energy of the virtual bound state
is given by the Anderson Coulomb matrix element with both electrons in
the one-electron orbital. For even weaker confinement in which the
length $L$ of the quantum dot exceeds the effective Bohr radius of the
electrons, the resonant bound states become strongly correlated with
the electron density tending to peak at opposite ends of the
dot-region in order to minimise their electrostatic repulsion
energy. This situation could occur in carbon nanotubes, for example,
where the Bohr radius is very small~\cite{gunlycke}. For such quantum
dots there is a low-lying singlet-triplet pair isolated from
higher-lying singlets and triplets. Thus we again have a single
singlet resonance and a single triplet resonance that dominates the
conductance near threshold.

In summary, within this simple formalism has been shown that quantum wires
with weak longitudinal confinement can give rise to spin-dependent,
Coulomb blockade resonances when a single electron is bound in the
confined region, a universal effect in one-dimensional systems with
very weak longitudinal confinement. The positions of the resulting
features at $G\sim \frac{1}{4}G_{0}$ and $G\sim \frac{3}{4}G_{0}$ are
a consequence of the singlet and triplet nature of the resonances.

\subsection{Thermopower}

The formalism applied above can be extended to include electrical and
heat currents through a region between two leads with different
temperatures and chemical potentials \cite{sivan86,proetto91}.  With
$T+\Delta T$, $\mu+eU$ for the left lead and $T$, $\mu$ for the right
lead, we get the electric current $j_e$ and the heat current $j_Q$
\begin{eqnarray}
I_{\mathrm e}&=&\frac{2e}{h}\int(\frac{1}{4}T_{0}+\frac{3}{4}T_{1})\Delta
f(\epsilon) {\mathrm d}\epsilon,\label{sest}\\
I_{\mathrm Q}&=&\frac{2}{h}\int(\epsilon-\mu)(\frac{1}{4}T_{0}+\frac{3}{4}T_{1})
\Delta f(\epsilon) {\mathrm d}\epsilon,
\end{eqnarray}
with
\begin{equation}
\Delta f(\epsilon)=f(\epsilon,\mu+eU,T+\Delta
T)-f(\epsilon,\mu,T).\label{deltaf}
\end{equation}

\begin{figure}[htb]
\center{\epsfig{file=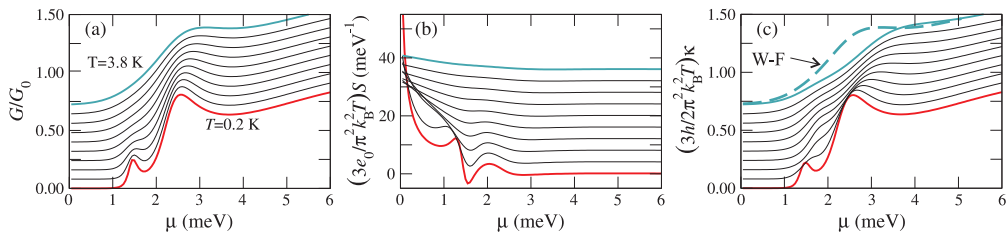,height=30mm,angle=-0}
\caption{ (a) Electrical conductance $G(\mu)$, (b) thermopower
$S(\mu)$, and (c) thermal conductance $\kappa(\mu)$ for circular wire
with parameters $a=10$~nm, $L=60$~nm, $v_0=12$~meV. Other parameters and
the numerical method is as in Ref.~\cite{rrj00} The dashed line in (c)
represents Wiedemann-Franz law result for $T=4$~K. The traces for
different $T$ are offset vertically for clarity.}\label{Fig5}}
\end{figure}

The Seebeck thermopower coefficient $S$ was for various systems
discussed in Refs.~\cite{proetto91,appleyard00}. It measures the
voltage difference needed to neutralise the current due to the
temperature difference between the leads. In the linear response
regime the thermopower is in our model given by \cite{thermo}
\begin{equation}
S(\mu)=\frac{U}{\Delta T}=-\frac{1}{eT}\frac{K_1(\mu)}{K_0(\mu)},
\label{cm}
\end{equation}
where
\begin{equation}
K_n(\mu)=-\int(\epsilon-\mu)^n(\frac{1}{4}T_{0}+\frac{3}{4}T_{1})
\frac{\partial f(\epsilon,\mu,T)}
{\partial \epsilon }
{\mathrm d}\epsilon.
\label{k}
\end{equation}
Eq.~(\ref{cm}) is formally the same as the Mott-Jones formula for
simple metals \cite{mott36} and generalized for a system with stronger
electron-phonon interactions in Refs.~\cite{jonson}.

In Fig.~\ref{Fig5}(a) the conductance is shown for a wire with a bulge
for different temperatures. The effect of temperature is accounted
only through the temperature of the Fermi function in
Eq.~(\ref{deltaf}).  Here the energy is measured from the threshold of
the conductance. In this case both, singlet and triplet resonances
contribute.  In Fig.~\ref{Fig5}(b) the thermopower is presented for
the same range of temperatures.  In the thermopower curve the dominant
structure at lower temperatures comes from the singlet resonance,
though the triplet resonance is still clearly discernible. At higher
temperatures the triplet structure is washed out first, in contrast to
the conductance result, Fig.~\ref{Fig5}(a).  The thermopower of
one-dimensional wires has been measured \cite{appleyard98,molenkamp90}
and more recently, further anomalies related to `0.7 anomaly' in
conductance were reported \cite{appleyard00}. The authors of
Ref.~\cite{appleyard00} observe a dip in $S(\mu)$ at energies
corresponding to the anomaly in $G(\mu)$.  However, the logarithmic
derivative with respect to the gate voltage of the measured $G$
exhibits a much deeper minimum than the dip in the measured $S(\mu)$,
which remains well above zero even at the lowest temperatures. This
clearly shows that a simple non-interacting formula is not valid in
this low temperature regime. Apart from the small corrections to the
logarithmic approximation to $S$, our results are in agreement with
the findings of Ref.~\cite{appleyard00}. That is, the calculated
thermopower is in good agreement with experiment except at low
temperatures where we also predict a deep minimum. This discrepancy at
low-temperatures may well be a many-body Kondo-like effect contained
within our model [Eq.~(\ref{Rejec_k})] but not within the two-electron
approximation we have used here.  At very low temperatures, a
Kondo-like resonance is expected \cite{boese01}, for which many-body
effects would dominate with a breakdown of formula Eq.~(\ref{cm}).

\subsection{Thermal conductance}

The linear thermal conductance is the heat current divided by the
temperature difference between the leads when the chemical potentials
are adjusted to give no electrical current. From Eqs.~(\ref{sest})-(\ref{k})
we see that this is related to the transmission probabilities by,
\begin{equation}
\kappa(\mu)=\frac{2}{h T}\left(K_2(\mu)-\frac{K_1^2(\mu)}{K_0(\mu)}\right),
\label{kapa}
\end{equation}
which for low temperatures simplifies to the Wiedemann-Franz law
\begin{equation}
\kappa(\mu)=\frac{\pi^2k_{\mathrm B}^2T}{3e^2}G(\mu).
\label{wf}
\end{equation}
In Fig.~\ref{Fig5}(c) $\kappa(\mu)$ is shown for $T$ from $0.2$~K to
$3.8$~K. A comparison of Fig.~\ref{Fig5}(a) with Figs.~\ref{Fig5}(c)
shows good agreement with the Wiedemann-Franz law at lower
temperatures but there is increasing deviation at higher temperatures
in the resonance region. For comparison, the dashed line in
Fig.~\ref{Fig5}(c) shows the corresponding linear approximation
result, i.e. the Wiedemann-Franz law.
One of the most striking features of these plots is that
$\kappa(\mu)$, calculated from Eq.~(\ref{kapa}), exhibits an anomaly at
higher energies than the corresponding anomaly in conductance, a
prediction which is open to experimental verification.

Within the framework of the Landauer-B{\" u}ttiker approach thermal
transport coefficients can be calculated for near-perfect
semiconductor quantum wires, which extends the work on spin-dependent
conduction anomalies. Thermopower and thermal conductance show
anomalies, related ultimately to the Coulomb interaction between a
localised electron and the remaining conduction electrons. We have
shown that the lower-energy singlet anomalies in thermopower are more
pronounced. These should be clearly observable in wires which show the
corresponding conductance anomalies, such as the narrow `hard
confined' wires reported in Ref.~\cite{kaufman99}, or in gated quantum
wires under high source-drain bias where the singlet anomaly is
clearly observed \cite{patel91}.

\subsection{Shot noise and 0.7 anomaly}

Recent high accuracy shot noise measurements enabled the
extraction of the Fano factor in ballistic quantum
wires~\cite{roche}. The Fano factor $F$ is a convenient measure of
the deviation from Poissonian shot noise. It is the ratio of the
actual shot noise and the Poisson noise that would be measured in
an independent-electron system~\cite{blanter}. This factor is, in
our model \cite{shot},
\begin{equation}
F=\frac{\int[T_{0}(1-T_{0})+3T_{1}(1-T_{1})](f_{L}-f_{R})^{2}{\mathrm
d} \epsilon}{\int(T_{0}+3T_{1})(f_{L}-f_{R})^{2}{\mathrm
d}\epsilon}.\label{eq:f}
\end{equation}
This expression and Eq.~(\ref{eq:lbg}), are based on the results
of a two-electron scattering between a single bound electron and a
propagating conduction electron with a summation over all
conduction electrons near the Fermi energy. This approximation is
only valid at temperatures above the Kondo scale in this
system~\cite{meir02}, as discussed in Ref.~\cite{rrj03}.
Eq.~(\ref{eq:f}) directly reflects the fact that singlet and
triplet modes do not mix in this pairwise interaction
approximation, resulting in contributions to the noise that add
incoherently with the probability ratio 1:3 for singlet and
triplet scattering.

\begin{figure}[htb]
\center{\epsfig{file=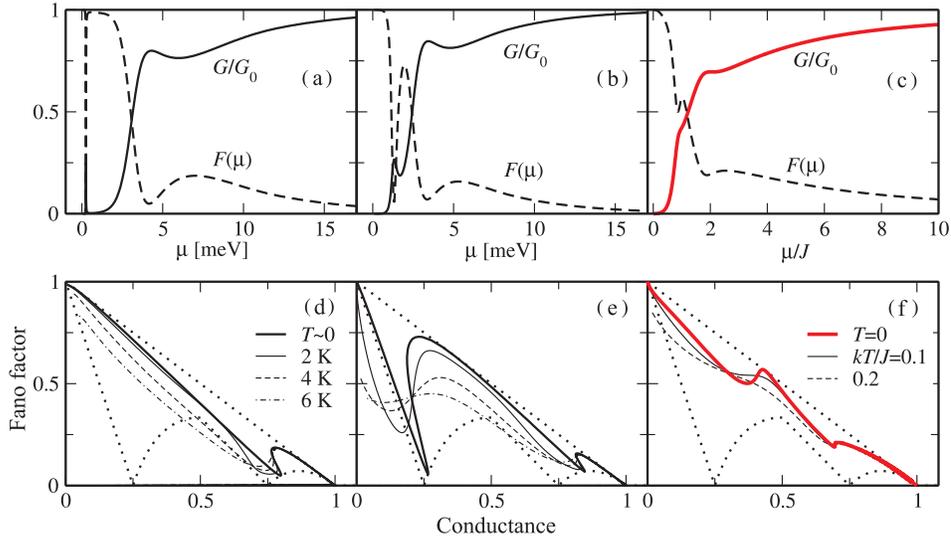,width=125mm,angle=-0}
\caption{\label{Fig6}Results for a cylindrical quantum wires with
a symmetric bulge for all parameters as in Fig.~3(b) and Fig.~3(c)
of Ref.~\cite{rrj00}. In panels (a) and (b) the conductance (full
lines) and the Fano factor are presented. Panels (d) and (e): Fano
factor as a function of $G$ (full line) and unitarity limits
(dotted line). The energy dependence of the Fano factor and
conductance for a left-right asymmetric confining potential (in
resonance $T_{S}<1$), (c), and Fano factor as a function of
conductance, (f).  Energy and temperature scale is here in the
units of single-triplet energy difference $J$.}}
\end{figure}

The conductance $G(\mu)$ and Fano factor $F(\mu)$ are plotted vs $\mu$
in Fig.~\ref{Fig6}(a) and Fig.~\ref{Fig6}(b), with Fig.~\ref{Fig6}(d)
and Fig.~\ref{Fig6}(e) showing the Fano factor $F$ vs $G$ for various
temperatures and in the linear response regime,
$\mu=\mu_{L}\sim\mu_{R}$. The dotted lines show zero temperature
boundaries for the allowed values of $F$, under the assumption of
validity of Eq.~(\ref{eq:f}) in the limit $T\to0$ and the unitarity
condition for the transmission probabilities, $0\leq
T_{S}\leq1$. Eq.~(\ref{eq:f}) is not strictly valid in the limit
$T\to0$ due to many-electron effects which start becoming important at
low temperatures. Thus this zero-temperature limit should be regarded
as a limiting behavior that would occur in the absence of such
many-body effects. The Fano factor exhibits two distinctive
features. Firstly, there is a structure for $G/G_{0}<0.5$
corresponding to the sharp $0.25$ singlet conductance anomaly. The
second distinctive feature is in the region $0.5<G/G_{0}<1$ and
corresponds to the dip in singlet channel just above the singlet
resonance and also partially to the triplet channel resonance. In our
previous work we assumed a symmetric confining potential fluctuation,
giving perfect transmission probabilities at resonance
energies. However, in real systems left-right symmetry will not be
perfect, especially if the fluctuation is of random origin, and also
under finite source-drain bias. In these cases $T_{S}<1$ even on
resonance. Such an example is presented in Figs.~\ref{Fig6}(c, f).  In
this case the structure of $F(G)$ is less pronounced, consisting of
kinks at $G/G_{0}\lesssim0.5$ and $G/G_{0}\lesssim0.75$. This behavior
is a consequence of the absence of a pronounced triplet resonance and
a dip in the singlet channel as mentioned above. Such a situation is
typical for very weak confining potential fluctuations, where the
triplet resonance is far in the continuum. The structure at
$G/G_{0}\lesssim0.5$ has the same origin as the ''0.25 structure'' in
conductance, a direct consequence of a sharp singlet resonance.

\section{Many-body effects}

Although the generic model we have used [Eq.~(\ref{hamilt2})] is
quite simplistic, it probably captures most of the essential
physics.  The many-electron generalisation of Eq.~(\ref{hamilt2})
may be obtained directly by expressing the Hamiltonian in
second-quantised form using the solutions of the one electron
problem, including both bound states in the well and unbound
states in the wire. The resulting Anderson-like Hamiltonian
\cite{anderson61} for the case of shallow well only, i.e. not
quantum-dot barriers, takes the form \cite{rejecand,rrj03},
\begin{eqnarray}
H &=&\sum_{k}\epsilon _{k}n_{k}+\epsilon _{d}n_{d}+ \sum_{k\sigma
}(V_{k}n_{d\bar{\sigma}}c_{k\sigma }^{\dagger }d_{\sigma }+
\mathrm{h.c.})+\nonumber\\
&+&Un_{d\uparrow }n_{d\downarrow } +\sum_{kk^{\prime
}\sigma }M_{kk^{\prime }}n_{d}c_{k\sigma }^{\dagger }c_{k^{\prime
}\sigma }+\sum_{kk^{\prime } }J_{kk^{\prime }}{\bf S }_{\bf d}\cdot
{\bf s}_{\bf kk^{\prime }}.
\label{Rejec_k}
\end{eqnarray}
Here $U$ is the Hubbard repulsion, $V_{k}$ is the assisted hopping
term, $M_{kk^{\prime }}$ corresponds to scattering of electrons and
the direct exchange coupling is $J_{kk^{\prime }}$. Spin operators in
Eq.~(\ref{Rejec_k}) are defined as ${\bf S}_{\bf d}=\frac{1
}{2}\sum_{\sigma \sigma ^{\prime }}d_{\sigma }^{\dagger }{
\hbox{\boldmath$\sigma$}}_{\sigma \sigma ^{\prime }}d_{\sigma ^{\prime
}}$ and ${\bf s}_{\bf kk^{\prime }}=\frac{1}{2}\sum_{\sigma \sigma
^{\prime }}c_{k\sigma }^{\dagger }{\hbox{\boldmath$\sigma$}}_{\sigma
\sigma ^{\prime }}c_{k^{\prime }\sigma ^{\prime }}$, where the
components of ${\hbox{\boldmath$\sigma$}}$ are the usual Pauli
matrices. A similar model has been proposed in Ref.~\cite{meir02}.
Although the Hamiltonian, Eq.~(\ref{Rejec_k}), is similar to the usual
Anderson Hamiltonian~\cite{anderson61}, we stress the important
difference that the $kd$-hybridization term above arises solely from
the Coulomb interaction, whereas in the usual Anderson case it comes
primarily from one-electron interactions.  These have been completely
eliminated above by solving the one-electron problem exactly. The
resulting hybridization term contains the factor $n_{d\bar{\sigma}}$,
and hence disappears when the localized orbital is unoccupied. This
reflects the fact that an effective double-barrier structure and
resonant bound state occurs via Coulomb repulsion only because of the
presence of a localized electron. The above Hamiltonian is difficult
to solve at low-energy/temperature, being in the class of many-body
Kondo-like problems. However, as with the Kondo problem, we expect
pairwise scattering solutions to capture the essential physics above
the effective energy/temperature where condensation phenomena becomes
important. It is this higher temperature scale that we have considered
in this paper and the results presented are entirely equivalent to
scattering solutions of the above Anderson-like model in the pairwise
scattering approximation. We must await further research to fully
elucidate low-temperature phenomena where it seems likely that the
Kondo effect, that is clearly observed in conduction through
quantum-dots, will be generalised to account for the suppression of
conductance anomalies in the low-temperature regime.~\cite{cron,meir02}

\section{Summary and conclusions}

In this paper we have mainly focused on a generic model of a
near-perfect quantum wire which contains a shallow quantum well
which behaves like an open quantum dot that captures a single
electron.  We have not investigated in detail the actual causes of
such a weak effective potential wells but point out that they may
well be due to quite different sources in different experiments,
e.g. thickness fluctuations, remote impurities or gates,
electronic polarisation, or some other more subtle electron
interaction effect. The main point is that because the effective
potential well is shallow, it will bind one and only one electron
and the subsequent spin-dependent scattering of a conduction
electron from this bound electron can be used to induce
entanglement between them. This is a consequence of resonant bound
states arising from the Coulomb interaction. When this is combined
with Pauli exclusion it gives rise to singlet and triplet
resonances and the generation of entanglement may be interpreted
as a filtering process in which either the singlet or triplet
component of the initial unentangled state is reflected with
complementary component being being transmitted. This same process
is also responsible for transport anomalies when many conduction
electrons (near the Fermi energy) scatter from a single bound
electron. This is again a filtering effect in which ~1/4 of
incident electrons are transmitted when the Fermi energy is close
the singlet resonance energy and ~3/4 are transmitted close to the
triplet resonance energy. The universal anomalies in conductance
and thermopower are a direct consequence of this and occur for a
wide range of circumstances in almost perfect quantum wires. This
universal behaviour also occurs in quantum dots with 'real'
barriers which will also exhibit singlet/triplet spin-filtering
and associated entanglement generation. However, for these cases
the resonances will be much sharper reflecting the well-known
Coulomb-blockade effect. If we imagine gradually increasing the
quantum-dot barriers from zero then the anomalies in conductance
will evolve into single-electron-tunneling resonances in the
coherent transport regime, thus providing a unified picture of
spin-dependent transport in quantum wire and quantum-dot
structures.

\end{document}